\documentstyle[11pt,paspconf,epsf]{article}
\def\refit{{\null}}
\def\etal{{\it et~al.\ }}

\def\simlt{\ {\raise-.5ex\hbox{$\buildrel<\over\sim$}}\ }
\def\kms{{km~s$^{-1}$}}
\def\ie{{\refit i.e.,}\ }

\def\etal{{\refit et al.\ }}

\def\refit{{\null}}
\def\subsun{_\odot}
\begin{document}

\title{Intracluster Stars and the Galactic Halos of Virgo}
\author{John J. Feldmeier\altaffilmark{1}, Robin Ciardullo\altaffilmark{1,2}, 
Roger Bartlett}
\affil{Department of Astronomy and Astrophysics, Penn State University,
525 Davey Lab, University Park, PA 16802}

\author{George H. Jacoby}
\affil{Kitt Peak National Observatory, National Optical Astronomy
Observatories, P.O. Box 26732, Tucson, AZ 85726}

\altaffiltext{1} {Visiting Astronomer, Kitt Peak National Optical Astronomy
Observatories, which is operated by the Association of Universities for Research in Astronomy, Inc., under cooperative agreement with the National Science Foundation.} 

\altaffiltext{2}{NSF Young Investigator}

\begin{abstract}

Planetary nebulae can be useful dynamical probes of elliptical galaxy halos, 
but progress thus far has been hampered by the small data samples 
available for study.  In order to obtain a large sample of planetary 
nebulae for dynamical analysis, we have re-observed the giant 
elliptical galaxy M87 in the center of the Virgo Cluster.  
Surprisingly, we find that the M87 sample is contaminated by a 
significant fraction of intracluster objects.  Follow up 
observations of three blank fields in the cluster confirm that 
a significant fraction of Virgo's starlight 
is intracluster.  We discuss the implications of 
intracluster stars for studies of galaxy halos in clusters.
\end{abstract}

\section{Planetary Nebulae as probes of galactic halos:}

Measuring the dark matter halos of elliptical galaxies 
is a difficult task.  Unlike disk galaxies, there are few 
dynamical tracers in ellipticals that can be observed over large
ranges in distance.  Stellar absorption features and emission line
gas disks cannot be used much beyond two effective radii 
($r_{e}$), and although X-ray mass determinations can be invaluable at 
large radii, currently X-ray observations cannot 
reach the centers of ellipticals, due to instrumental resolution, 
and potential complications due to cooling flows. 
  
Planetary nebulae (PN) are objects that are extremely well suited to bridge
the gap between inner and outer dynamical methods.  Individual PN can 
be easily detected at several $r_{e}$ with 4-m class telescopes out to 
distances of $\sim 20$~Mpc., and samples already exist for over 
twenty elliptical galaxies.  Since PN emit a large fraction ($\sim 15\%$) 
of their flux in a single narrow emission line at 5007~\AA, 
their velocities can be determined with a short
duration spectrum at moderate, $\lambda /\Delta \lambda \sim 5000$
resolution.  Finally, PN are excellent distance indicators (for a review,
see Jacoby 1996a), 
through the Planetary Nebulae Luminosity Function distance 
method (PNLF), which allows the true physical scale of the 
system to be determined.

\begin{figure}
\caption{This is our $16' \times 16'$ [O~III] image of M87, created
by combining seven one-hour exposures with the Kitt Peak 4-m 
telescope.  North is up, and east is to the left.  The 337 PN candidates 
found in this survey are denoted as squares.}
\end{figure}

There have been a handful of detailed dynamical studies of 
elliptical galaxies out to several $r_{e}$, using PN 
(for a current review, see Arnaboldi \& 
Freeman 1997).  However, with the notable exception of 
NGC~5128 (Hui \etal 1995), 
these studies suffer from relatively poor statistics;  typically, 
the samples observed to date are less than a hundred objects.  
Without more data, it is impossible to explore
multi-component dynamical models of halos: the only analysis possible 
is to assume simple models and test for consistency.  Clearly, this could
lead to erroneous results.  For example, if elliptical galaxies are
formed via encounters between galaxies, then box or tube orbits along
the galaxy's major or minor axis may be the rule, and most of the
system's angular momentum (and possibly mass) may reside far in the
halo (Barnes 1988, 1992).  If this is the case, then the stellar
distribution function will be extremely anisotropic and the
mass-to-light ratio will change with radius.  To investigate these
possibilities, several hundred test particle velocities
(Merrifield \& Kent 1990; Merritt \& Saha 1993) are needed.

To rectify this limitation, we re-observed the giant elliptical 
galaxy M87 with a wide field ($16' \times 16'$) CCD camera on the
Kitt Peak 4-m telescope.  This survey 
yielded a total of 337 PN candidates for use in a dynamical study.

\section{The Problem: an unusual luminosity function}

\begin{figure}
\plotfiddle{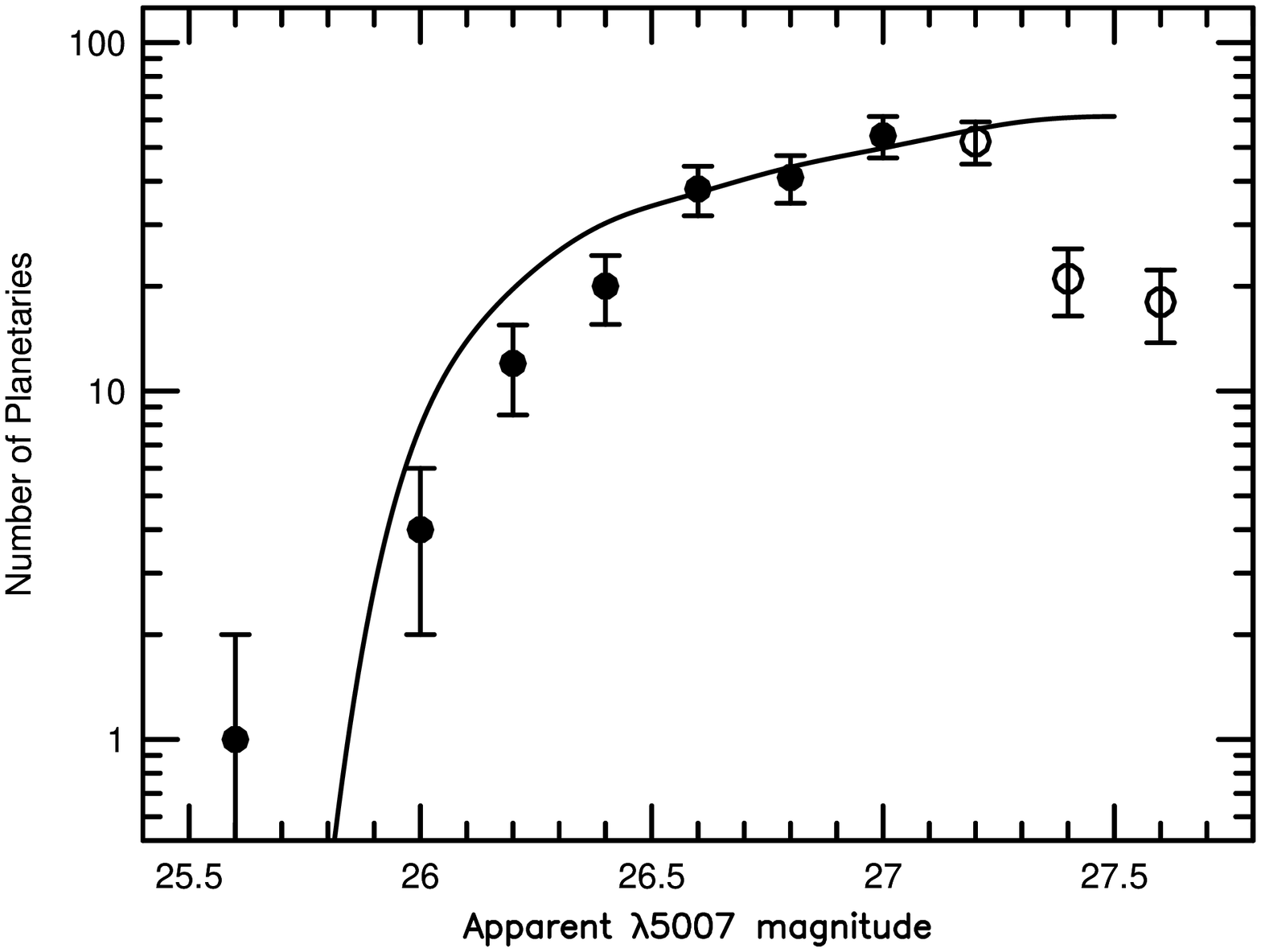}{80pt}{-90}{33}{33}{-199}{149}
\plotfiddle{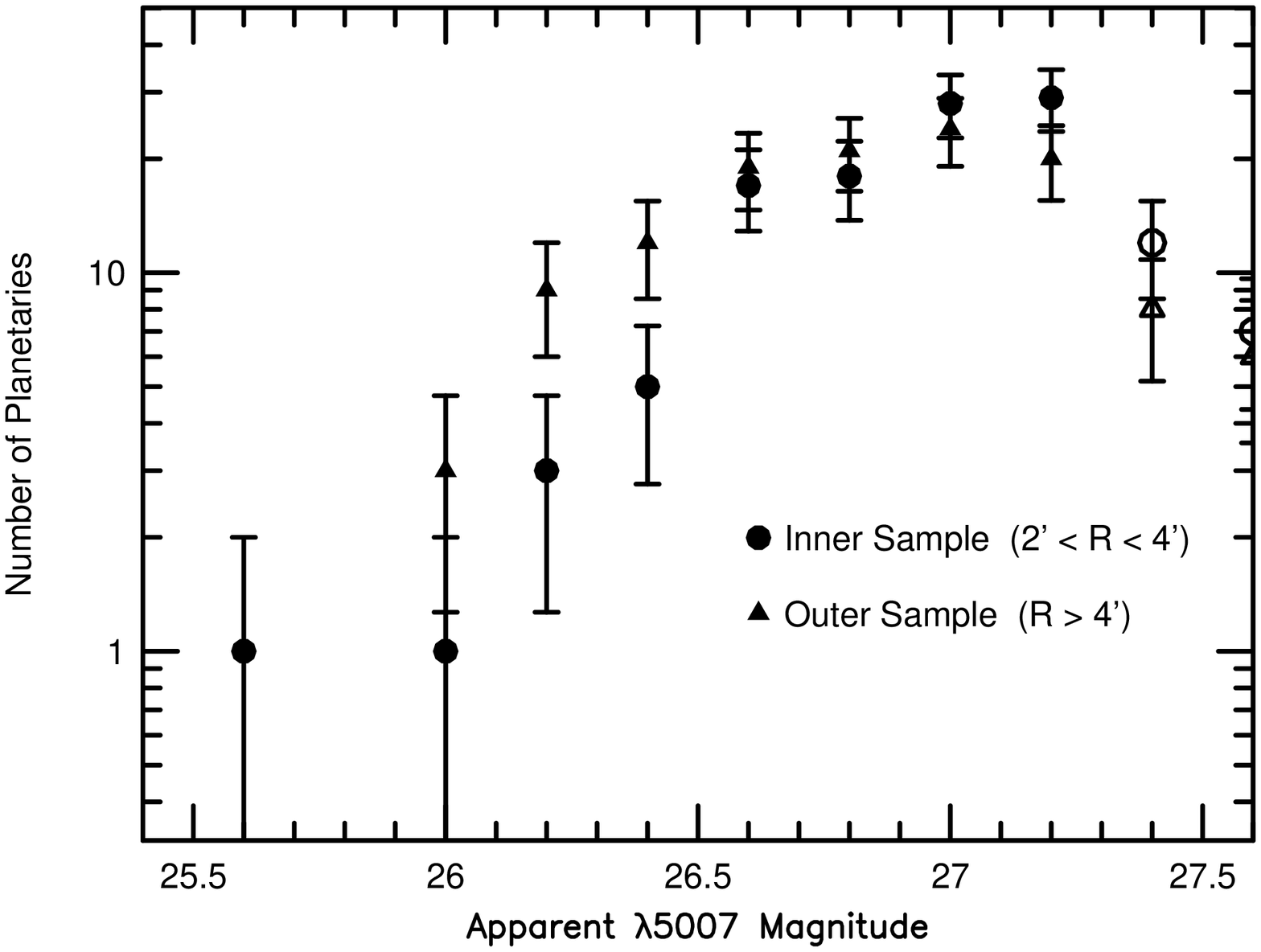}{80pt}{0}{33}{33}{1}{-19}
\caption{On the left is the planetary nebula luminosity function for a 
statistical sample of M87 planetaries.  The curve shows the empirical
function shifted to the most likely distance modulus.  Although this
curve is the ``best fit'' to the data, it is still excluded at the
99.9\% confidence level.  On the right, we have divided the sample of
M87's PN into two samples, as described in the text.  As can be clearly seen
the outer sample has many more bright objects.}
\end{figure}

Figure~2 (left) plots the planetary nebula luminosity function 
for a statistical sample of M87's PN{}.  
From these data, the PNLF distance to the galaxy can
normally be derived by convolving the empirical model for the PNLF given
by Ciardullo \etal (1989):
\begin{equation}
N(M) \propto e^{0.307M} \, [1 - e^{3(M^{*}-M)}]
\end{equation}
with the photometric error function and fitting the data to the 
resultant curve via the method of maximum likelihood.  However, in the
case of M87, the most likely empirical curve (the solid line) is a 
poor fit to the luminosity function.  Kolmogorov-Smirnov 
and $\chi^2$ tests both exclude the Ciardullo \etal 
law at the 99.9\% confidence interval.  This is a remarkable result: 
none of the PNLFs from any of the $\sim 30$ previously 
studied galaxies differs significantly from the empirical law.  Moreover, 
the large number of PN detected in this survey cannot be invoked to explain 
the discrepancy.  The luminosity functions of M31, M81, NGC~5128, 
and NGC~4494 all contain similar numbers of objects.  The planetaries 
surrounding M87 are therefore unique in some way.

An even more surprising result comes if we divide our PN sample 
in two, and compare the PNLFs of M87's inner and outer halo (first noted
in Jacoby, 1996a).  For the inner sample, we take all the PN 
in our statistical sample 
with isophotal radii between $2\arcmin$ and $4\arcmin$; for the outer
sample, we take those PN with $r_{\rm iso} > 4\arcmin$.  
Both samples are plotted in Figure~2 (right).  Of the 20 brightest PN, 18 
belong to the outer sample.  More importantly, the shapes of the 
two PNLFs appear different: a Kolmogorov-Smirnov test reveals that the two 
samples are different at the 92\% confidence level.  Again, this result is 
unique.  Explicit tests for changes in the PNLF cutoff with galactocentric 
radius have been performed with the large samples of PN available in NGC~5128 
and NGC~4494.  In neither case was a gradient observed.  

\section{The Explanation:}

Internal and external tests on the $\sim 30$ early and late-type galaxies 
surveyed to date have shown that the location of the PNLF cutoff 
does not depend on the properties of the parent stellar population 
(cf.~Jacoby 1996b).  However, a number
of mechanisms do exist which can, at least in theory, cause the PNLF
technique to fail and produce a change in the observed value of $m^*$.  
The first, and simplest, is to hypothesize that some instrumental effect 
exists, such as a radial gradient in the flatfield or the transmission curve 
of the filter.  We have run a number of tests, and have ruled out an
instrumental effect in our data.  

A second possibility is to invoke 
non-uniform extinction in M87.  Dust has been detected in the central
regions of many elliptical galaxies (Ebneter, Djorgovski, \& Davis 1988), 
and it is possible in principle that dust might cause the difference 
between our inner and outer samples.  However, studies of dust have 
concentrated on the central regions of elliptical galaxies,
while our survey deals exclusively with PN that are more than 1.5 $r_e$ 
from the galactic nucleus.  Moreover, even if there is 
a strong gradient in the dust distribution, this still 
may not translate into an observed gradient in $m^*$.  As Feldmeier, 
Ciardullo, \& Jacoby (1997) have shown, the location
of a galaxy's PNLF cutoff is relatively insensitive to the presence of
dust, as long as the scale length for the obscuration is smaller than that
of the stars.  This makes it very unlikely that dust is responsible 
for the change in $m^*$.
  
Although the absolute magnitude of the PNLF cutoff is extremely 
insensitive to the details of the underlying stellar population, a dramatic
change in the metallicity or age of M87's halo stars could, in principle,
produce a change in $m^*$ similar to that observed.  Both observations 
and theory suggest that galaxies with [O/H] $\simlt -0.5$ can have 
a PNLF cutoff that is different from that of metal-rich populations by 
$\sim 0.1$~mag.  Unfortunately, this effect acts in the wrong direction:
it is the metal-poor systems that have fainter values of $M^*$.  In 
order to explain the observed gradient, the center of M87 would have to
be metal-poor, and the halo would need to be metal-rich.  Observations
show that this is extremely unlikely (Kormendy \& Djorgovski 1989).

Similarly, it is difficult to use population age to explain PNLF variations.
According to the models of Dopita, Jacoby, \& Vassiliadis (1992) and M\'endez 
\etal (1993), the location of the PNLF cutoff is nearly independent of age
for populations between 3 and 12~Gyr.  If these models are correct, then in
order to enhance the luminosity of the PNLF cutoff in M87's outer halo, one
must hypothesize an unrealistically young age for the stars, $\sim 0.5$~Gyr.  

The best hope for explaining the observed changes in M87's halo PNLF lies
in the Virgo Cluster itself.  All of the above explanations implicitly
assume that the PN projected onto M87's outer halo are at the same
distance as those PN which are members of the inner sample.  However, if the 
Virgo Cluster has a substantial population of intracluster stars, this will 
not be the case, as some objects will be superposed in the foreground, and
others will be in the background.  For example, if the Virgo Cluster is at a
distance of $\sim 15$~Mpc, then the central $6^\circ$ core of the cluster
(de Vaucouleurs 1961), has a linear extent of $\sim 1.5$~Mpc.  If the core
is spherically symmetric and filled with stars, then we might expect some 
intracluster objects to be up to $\sim 0.25$~mag brighter than the value of 
$m^*$ derived from galaxies at the center of the cluster.  This is roughly
what is observed in Figure~2.

Further evidence that the anomalous PNLF of M87 is due to foreground 
contamination comes from the fact that it is the outer sample of objects
that has most of the bright PN{}.  The number of foreground PN detected 
in any region of our CCD field should be roughly proportional to the area 
of the field; since the outer region samples $\sim 8$ times more area than 
the inner field, those data should contain $\sim 8$ times more intracluster
objects.  In addition, M87's sharply peaked surface brightness profile
guarantees that the ratio of galaxy light to intracluster light in the
inner field will be much larger than that in the outer field.  Consequently,
the contribution of intracluster objects to the inner sample will be small,
while that for the outer sample will be relatively large.  Again, this is
roughly what is displayed in Figure~2.

Although the idea of intracluster planetary nebulae (IPN) seems 
speculative, there is, in fact, conclusive evidence that 
such objects do exist.   In their
radial velocity survey of 19~PN in the halo of the Virgo Cluster elliptical
M86 ($v = -220$~\kms), Arnaboldi \etal (1996) found three objects with
$v > 1300$~\kms.  These planetaries are undoubtably intracluster in origin.  
Significantly, the PN observed by Arnaboldi \etal were originally identified 
with a 30~\AA\ filter centered at 5007~\AA\ (Jacoby, Ciardullo, \& Ford
1990); intracluster objects with $v > 1000$~\kms\ should have been strongly
excluded.  The only reason three were detected was that at $v \sim 1500$~\kms,
[O~III] $\lambda 4959$ is shifted into the bandpass of the [O~III] $\lambda
5007$ filter!  Since $I(\lambda 5007)/I(\lambda 4959) = 3$, only the very
brightest PN could have been detected in this way.  The existence of three
intracluster objects in the Arnaboldi \etal sample therefore implies the
presence of many more.  

\section{The Confirmation:}
Motivated by our studies of M87, we decided to search the Virgo 
Cluster to confirm the presence of intracluster PN.  
We observed three $16' \times 16'$ blank fields: one at the 
isopleth center of subclump A (subclumps defined by Binggeli, 
Tammann \& Sandage 1987) 52' away from M87, one $\sim 36\arcmin$ 
north of the giant 
elliptical NGC~4472 in subclump~B, and a third due
north of the M87 field shown above.  In the first two fields, we 
identify 69 and 16 intracluster planetary nebulae candidates, 
respectively.  In the third field near M87, we detect 75 planetary 
nebulae candidates, of which the majority can be attributed to M87.
However, like the M87 data shown above, the empirical PNLF is again a poor fit
to the observed luminosity function observed for Field 3: a 
Kolmogorov-Smirnov test excludes the Ciardullo \etal (1989) law at 
the 93\% confidence level.  From our
determination of the distance to M87 (Ciardullo \etal 1997), and the
expected PNLF for M87 PN, we can show that at least ten planetaries, 
and probably more, are intracluster in origin.  The combined data from 
the three fields confirms the intracluster hypothesis for M87.  

\begin{figure}
\plotfiddle{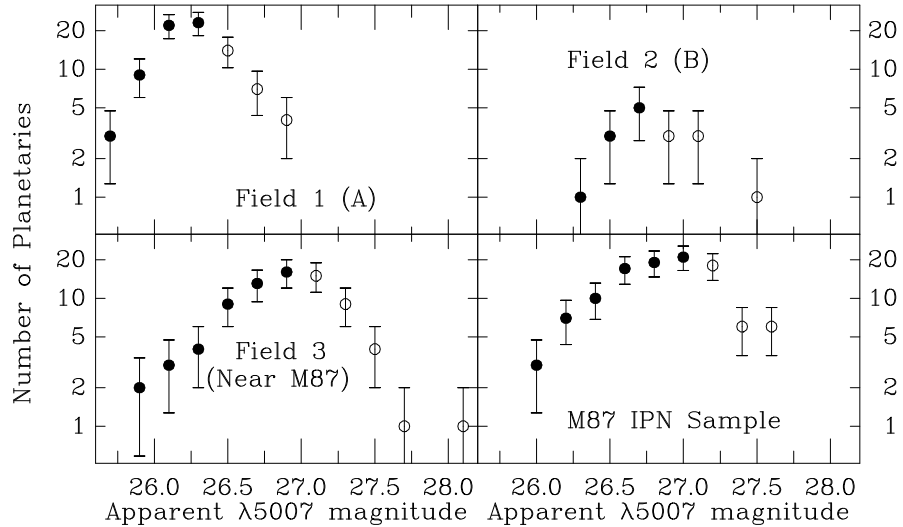}{200pt}{0}{60}{60}{-175}{-227}
\caption{The planetary nebula luminosity functions for the three intracluster
fields, plus a sample of suspected intracluster planetaries from M87, 
binned into 0.2~mag intervals.  The solid circles represent 
objects in our statistical IPN samples; the open circles indicate 
objects fainter than the completeness limit.  Note that the numbers 
of intracluster planetaries vary from field to field.}
\end{figure}

\section{PN as probes of intracluster starlight: some preliminary results}

Intracluster PN are interesting objects in their own right, not simply
contaminants to galactic halo samples.  Ever since Zwicky 
(1951) first claimed the detection of excess light between galaxies 
of the Coma Cluster, intracluster light has been of great 
interest to astronomers.  Depending on the efficiency of 
tidal stripping, this intracluster component will contain 
anywhere between 10\% and 70\% of the cluster's total 
luminosity (Richstone \& Malumuth 1983; Miller 1983).  It 
is therefore a sensitive probe of how tidal-stripping works, 
the distribution of dark matter around galaxies, and of the 
initial conditions of cluster formation.

We can use our PN observations to trace the amount of
stellar light distributed in the Virgo intracluster medium.  
Renzini \& Buzzoni (1986) have shown that the bolometric-luminosity specific 
stellar evolutionary flux of non-star-forming stellar populations, should 
be $\sim 2 \times 10^{-11}$~stars-yr$^{-1}$-$L_{\odot}^{-1}$, independent of 
population age or initial mass function.  If the lifetime of the planetary 
nebula stage is $\sim 25,000$~yr, then every stellar population should have 
$\alpha \sim 50 \times 10^{-8}$~PN-$L_{\odot}^{-1}$.  Observations in 
ellipticals have shown that no galaxy has a value of $\alpha$ greater than
this number, though $\alpha$ can be a factor of five smaller 
(Ciardullo 1995). Nevertheless, the direct 
relation between the number of planetary nebulae and the parent 
system's luminosity allows us to estimate the number of stars in 
Virgo's intergalactic environment.  When we apply this calculation
to our data, taking the $\alpha$ value that produces
the minimum amount of starlight necessary to reproduce the data, we find 
an estimated $6 \times 10^{9} L\subsun$ in the two $16' \times 16'$ 
intracluster fields. 
We note that this is the minimum amount of starlight necessary to 
explain the data, and the amount could
be greater by up to a factor of five.  If we compare this to the 
amount of starlight in galaxies, we find that the intracluster 
stars produce at least 26\% of the total stellar luminosity of Virgo.  

Another interesting result is the dramatically different
numbers of intracluster PN in Fields 1 and 2.  Although the 
survey depths of the regions differ (due to differences in sky 
transparency and seeing), it is clear that the highest density of 
intracluster objects is in Field~1, at the center of subclump~A{}.  
After accounting for the differing depths, 
the density of PN in Field~2, which is near the edge of the $6^\circ$ 
Virgo Cluster core in subclump~B, is down by a factor of $\sim 4$. 
The fact that subclump~B has fewer PN than subclump~A can probably
be attributed to cluster environment.  It is well known that subclump~B
has fewer early-type galaxies than subclump~A (Binggeli, Tammann, 
\& Sandage 1987).  If ellipticals and intracluster stars 
have a related formation mechanism (\ie galaxy interactions), then 
a direct correlation between galaxy type and stellar density in 
the intergalactic environment might be expected.  

Clearly, further data is needed to confirm and support these results.  
However, it is clear that there is a significant amount of intracluster
starlight in the Virgo Cluster that must be taken into account when
dynamical studies are undertaken.  If intracluster objects are ignored, 
then the derived M/L ratio will be overestimated, due to the much larger
velocity dispersion associated with the cluster.  

\section{Future plans and prospects:}

The analysis of the dynamics of Virgo's stars has already begun.
We have obtained data on M87's PN velocities from the 
WIYN telescope, and will soon begin observing 
intracluster PN with the Hobby-Eberly Telescope (HET).  
Although our sample of M87 objects is contaminated with 
intracluster objects, it may be possible to separate out the 
populations using the PNLF and by modeling the velocity distribution
of intracluster objects.

In order to learn much more about intracluster PN, we are continuing our 
[O~III] $\lambda 5007$ Virgo survey.  Intracluster PN provide 
information on both the two dimensional and, by using 
the PNLF, three dimensional structure of the cluster.  
Through the velocities and spatial distribution of intracluster PN, we can 
determine whether models of ``galaxy harassment'' (Moore \etal 1996) 
can explain the large amounts of intracluster starlight.
In the future, intracluster stars should be a useful tool in understanding
galaxy clusters, and the nature of galactic halos.

\acknowledgments

We would like to thank the organizing committee for giving us the opportunity 
to present this work orally.  This work was supported in part by 
NASA grant NAGW-3159

\end{document}